# An Efficient Mobile Gateway Selection and Discovery based-Routing Protocol in Heterogeneous LTE-VANET Networks

Driss Abada[1], Rachid Adrdor[2], Omar Boutkhoum[1], and Adil Bohouch[3]

[1]Department of Computer Sciences, Faculty of sciences, Chouaïb Doukkali University, El Jadida, Morocco
[2]Department of Computer Sciences, Faculty of sciences, Ibn Zohr University, Agadir, Morocco
[3]Department of Educational Technologies, Faculty of sciences of education, Mohamed V University, Rabat, Morocco

## ABSTRACT

*Coupling cellular communication networks with vehicular ad hoc networks (VANET) can be a very interesting way out for providing Internet access to vehicles in the road. However, due to the several specific characteristics of VANETs, making an efficient multi-hop routing from vehicular sources to the Internet gateways through Long Term Evolution (LTE) technology is still challenging. In this paper, an Internet mobile gateway selection scheme is proposed to elect more potential vehicles to behave as gateways to Internet in VANETs. Therefore, the discovery and the selection of route to those mobiles gateways is carried out via an efficient multiple metrics-based relay selection mechanism. The objective is to select the more reliable route to the mobile gateways, by reducing the communication overhead and performing seamless handover. The proposed protocol is compared with one recent protocol based on packet delivery ratio, average end-to-end delay and overhead. The results show that the proposed protocol ameliorates significantly the network performance in the contrast of the other protocol.*

## 1. INTRODUCTION

Internet access for VANETs will be an integral part of transport intelligent system applications. Indeed, this intelligent transportation field is constantly evolving, especially in recent years where many wireless technologies have been applied to effectively share and provide data services. Available technologies include, IEEE 802.11p [1], [2], UMTS (3G) [3], LTE and LTE-Advanced (4G) [1],[4] and recently millimeter wave vehicular communications (5G) [5].

VANETs are a promising technology that has much potential in the development of Intelligent Transport Systems (ITS) by enabling a wide range of applications and services on the road. In these networks, high-speed mobile units, in particular vehicles, tend to communicate with each other, or even with roadside infrastructure. The WAVE (Wireless Access for Vehicular Environments) protocol stack specifies an architecture including a IEEE 1609 set of standards that are integrated with the IEEE 802.11p/WAVE standard to adapt existing technologies (802.11), support new technologies (such as IPv6), and define security mechanisms.

IEEE 802.11p is the standard dedicated to vehicle-to-vehicle (V2V) and vehicle-to-infrastructure (V2I) communications in the DSRC band, on which the WAVE stack is based for the support and the improvement of the physical layer and the media access control (MAC) layer. The IEEE 802.11p band is 5.9 GHz, consisting of seven channels, each with a frequency equal to 10 MHz, and can communicate within a short radio range of about 300 m. It allows data to be exchanged at a rate of up to 27 Mbps. IEEE 802.11p protocol has shown its effectiveness for communicating vehicles within a VANET network, but many researchers [6],[7] have demonstrated that in many situations due to different characteristics of VANETs such as high mobility and unexpected driver behavior, there are still several issues such as frequent link disconnection, network topology changes, frequent fragmentation, reliability requirement of safety applications, limited coverage and so on, that make IEEE 802.11p behave inefficiently and disable to satisfy quality of services requirement of multiple application of VANETs and different requesting communication requirement of ITS services. With the aim of overcoming certain constraints, various works have been proposed to integrate VANETs and cellular communication networks as complementary components in a heterogeneous network architecture. IEEE 802.11p offers high data rate while cellular networks provide extensive coverage and support mobility. The General Packet Radio Service (GPRS), Universal Mobile Telecommunications System (UMTS) and Long Term Evolution (LTE) are remarkable ones.

LTE is a wireless access technology, categorized as 4G wireless broadband. It was developed under the Third Generation Partnership Project (3GPP). The LTE operates in a packet-switched manner and does not have a circuit-switched domain, unlike GSM/GPRS and UMTS [8]. LTE is intended to manage better system capacity and wireless coverage over large geographic areas. It brings many advantages including support for multiple antennas, seamless incorporation with other systems, lower latency, higher peak data rates, operating costs, etc. Theoretically, LTE is 5 to 10 times faster than 3G(UMTS) since it provides high-speed data download measured in several hundred megabits per second (Mbps), compared to some tens of Mbps for 3G and it can offer transfer latency lower than 5ms [6]. The LTE technology effectiveness, either in cost or in performance, is due to its simplified network architecture which is based on Internet Protocol (IP) and to advanced algorithms which allow better use of resources. In LTE, the base station eNBs (evolved NodeBs) provide the gateway between mobile terminals, radio antennas, and the core network of LTE operators to optimize all the UE's (User Equipment) communication in a flat radio network structure. The eNB has many features within an LTE network such as Radio Resource Management, Routing data from the user plane to the service gateway, planning and transmission of broadcast information, etc.

According to [9], utilizing protocols in heterogeneous networks instead of solely relying on IEEE 802.11p can lead to improved reliability in vehicular communication and significantly reduce undesired delays in various vehicular applications. This paper introduces a novel protocol called the Internet mobile gateway selection and discovery based-routing protocol (IMGsdRP) that aims to provide a reliable and efficient method for establishing Internet access in VANETs. The protocol is specifically designed to maintain a stable and long-lasting connection to the Internet while ensuring high throughput and meeting the quality of service requirements of applications in the ITS environment. Overall, the IMGsdRP protocol offers a promising solution for addressing the challenges associated with Internet connectivity in VANETs. The innovative IMGsdRP routing protocol has been meticulously designed to confront the aforementioned challenges and accomplish the stated objectives through a series of significant contributions, as enumerated below.

- By integrating key metrics to carefully select potential Internet mobile gateways and establish reliable multi-hop pathways to them, IMGsdRP protocol can satisfy the quality of services, particularly with respect to throughput, latency, and data packet delivery for Internet access applications in VANETs, ensuring seamless and uninterrupted connectivity for vehicles in the road.

- The IMGsdRP protocol can make efficient use of the limited network resources, particularly bandwidth. For this purpose, the protocol used timer-based selection strategy to select the most suitable vehicles to act as gateways and then eventually opt for relays to achieve their IEEE 802.11p coverage without flooding network with unnecessary traffic.

- The protocol IMGsdRP adopted a hybrid mobile gateway discovery process where it combined the advantages of proactive and reactive approaches. The objective is to reduce delay and overhead associated with establishing a route to the mobile gateways. Moreover, possessing a group of routes to various mobile gateways and their corresponding lifetime in its routing table, a source vehicular node can seamlessly transfer the connection to the new Internet gateway before the current one ends. Thus, the IGMsdRP can perform a seamless handover at a reasonable time.

- The IMGsdRP protocol utilizes some particular vehicles, such as public transport vehicles like taxis, and buses, as mobile gateways. This approach intercepts continual handoffs at LTE base stations and reduces the associated signaling overhead by only having a limited number of mobile gateways present in the network.

The layout of this paper is as follows. Section 2 presents the review of the existing related works. The proposed routing protocol IMGsdRP is detailed in Section 3. The performance of the proposed mechanisms is evaluated in Section 4. Finally, we give the conclusion and future work directions in Section 5.

## 2. RELATED WORKS

Several contributions in the open literature have studied extensively the problem of mobile gateway selection and discovery in the mechanism of multi-hop relay for connecting VANETs to the Internet in an integrated heterogeneous wireless network.

In the paper [9], the authors have developed a robust and trust mobile gateway selection (RTMGwS) protocol capable of constantly connecting to the Internet and providing other services while traveling on the road. This selection is based on some particular parameters which reflect the VANETs characteristics and the degree of trust of vehicles. The proactive mechanism of establishment and maintenance of route to the mobile gateways is similar to that of the work [10] proposed to connect VANET to the Internet via fixed infrastructures, whereas, the reactive mechanism is alike to that in [3]. Thus, in proactive mobile gateway discovery method, the protocol has based on multi-hop relay selection to inform source vehicles of the available mobile gateway by receiving an advertised message transmitted periodically. For designating vehicles to act as mobile gateways in the network, the protocol used received signal strength quality metric. The reactive method is executed in two main cases: (1) a vehicular source does not hear any announcement messages, and (2) sundry vehicles in the BST communication range have activated their EUTRAN interfaces. In this protocol, the authors assumed that all vehicles have dual interfaces and can act as gateways at any moment if they are inside UMTS BST base station coverage and satisfied the received signal condition. However, if all vehicles can transmit directly via the UTRAN UMTS interface, the UMTS BST base station will be saturated due to more MAC frames exchanging between vehicles and the UMTS BST base station. Consequently, excessive consumption of cellular network resources such as bandwidth, especially in a dense vehicle environment, will occur. In the present work, between several neighboring gateways one potential vehicle will be designed to act as gateway, which will lead to minimizing the number of vehicles acting as gateways in UMTS BST coverage.

In [11], an enhanced hybrid wireless mesh protocol (E-HWMP) protocol was proposed to form multi-hop routing through VANETs, to allow vehicles equipped with 802.11p/802.11s to connect to the Internet via LTE eNodeB. In this protocol, a new scheme of clustering for the multi-hop relay in vehicular communication is implemented. The three metrics received LTE signal strength, available route capacity, and stability represent the main elements used to select the gateway and to outrank the optimal gateway node. To calculate the weight of these metrics, the proposed scheme is based on the Simple Additive Weighting (SAW) technique. In addition, the protocol adopted the hybrid approach that generally mixes proactive and reactive modes. The main goals of this approach are to minimize the delay caused by the route discovery in a reactive approach and to reduce overhead generated by periodic broadcasting in a proactive procedure. However, adding to the delayed discovery of paths, the disadvantage of the reactive method used in this work is that the load is increased on the relay vehicles, especially those located in the transmission range of the gateway. In addition, in this protocol, the selection of the gateway is carried out on the side of the source node. That means that all gateways that receive route request messages will send a replay. This results in an increase in the number of control messages exchanged in the network. In the current work, a hybrid method in which the proactive/reactive approaches of routing are combined to minimize their disadvantages. In both approaches, our protocol bases on distributed timer strategy to select adequately forwarding vehicles with minimum broadcasts.

In this paper [3], the author introduced a new architecture combining both 802.11p-based-VANET and 3G/UMTS wireless technologies. This architecture aims to allow vehicular nodes to communicate efficiently with 3G/UMTS base stations by dynamically clustering gateway vehicles based on movement direction, UMTS-RSS, and IEEE 802.11p transmission ranges. If the mobile gateway was nearest to the center of the cluster, then the cluster leader was designated in a cluster. Moreover, the author proposed an adaptive mechanism for mobile gateway management allowing VANETs to have access to the UMTS network. The mechanism aims to make all vehicles able to send and/or receive data from the Internet via the UMTS network by electing a few potential vehicles to play the role of an Internet gateway. This approach prevents the saturation of base stations, making communication with them limited to certain vehicles only. Clustering can attenuate the effect of redundancy control messages by helping vehicles better manage inter-vehicle communications. Allowing multiple gateways to operate at once can help to eliminate bottlenecks and congestion in the network. The clustering is an efficient mechanism to reduce significantly the control message overhead and to make reliable communication between vehicular nodes in VANETs. However, the active clustering technique adopted in this work is based on periodic beacon packets to form and maintain clusters, thus clustering process and the cluster head selection are complex and may require more signaling overhead and extra time than those dedicated to data traffic exchange. Additionally, the best solution may not be to select the cluster head as the nearest vehicle to the center point of the cluster because this vehicle, depending on its speed, may leave this point quickly. In our protocol, the vehicular gateway selection scheme may consider as one hop passive clustering of VGCs which operates without any extra-message and the cluster head is selected without flooding the network. The cluster head is the node selected to act as Internet gateway in our protocol.

The authors in [12] has suggested a new Simplified Gateway Selection Scheme (SGSS) to enlarge the communication area of the VANET network and reduce the recurrent handover action allowing vehicles to maintain connection long time to the UMTS backbone network. The best routes to the mobile gateways are selected using three parameters, namely, route stability, available route capacity, and UMTS received signal strength. The proposed scheme is integrated into the existing DSDV and AODV mobile ad hoc routing protocol. The SGSS aims to extend the network communication zone by coupling of VANETs with the UMTS network. Since the protocol had not integrated any metric to choose the lower hop-count route, if the zone of VANET adopted is large, somewhat source vehicles can transmit data via many numbers of hops. As a consequence, the protocol may suffer from high packet delays. Moreover, the proposed scheme used metrics to prevent bottleneck problems on the gateway and intermediate node. Even this, one node can be overloaded during the traffic exchange. It will be necessary to define an explicit mechanism to avoid as much as possible bottlenecks in the network.

In the research work [10], the authors introduced a routing protocol allowing Internet access for VANETs using static gateways placed along the road. The selection of a route is performed based on two metrics such as link expiration time (LET) and route expiration time (RET). Thanks to these metrics, it can proactively disseminate announcement messages across multiple hops, eliminating network inundation, can seamlessly perform hand-overs, and can select efficiently the most stable routes to these stationary units. However, IEEE 802.11p-based-Internet access for VANETs is not more efficient due to various encountered issues, which are cited above. In addition, to ensure their serviceability, these protocols require the installation of many roadside units (RSUs) along the road. In this paper, we assume that public transport vehicles will be employed to behave as Internet gateways [12], instead of RSU. In contrast, RSUs can be applicable for security and traffic applications of VANETs such as fraud detection, jam detection, weather forecasting, etc.

## 3. PROPOSED ROUTING PROTOCOL DESCRIPTION

### 3.1. Proposed IMGsdRP Architecture Model

The designed VANET-LTE heterogeneous network architecture is shown in Figure 1. The network area is a highway where the traffic flows in both directions, with two lanes for each direction. The architecture includes a LTE Evolved Node B (eNodeB) base station transceiver that is deployed on the roadside. All vehicular nodes in this architecture are equipped with IEEE 802.11p radio interfaces that is enabled and activated all time to allow them to communicate between each other in V2V mode, whereas, in addition, some of these vehicles have a second interface which is Evolved Universal Terrestrial Radio Access Network (EUTRAN) interface, for enabling them to exchange data with the eNodeB, to access the essential elements of the LTE network. For geolocation, we suppose that all vehicles are equipped with Global Positioning System (GPS), which will allow them to track their movements.

The principal aim of this architecture is to discover vehicular Internet gateways equipped with dual interfaces, which can assist other vehicles having only 802.11p wireless network interface or those outside the interest area, to access to the Internet. In the report of this architecture, a single region of vehicular networks is covered by the BST, where the received signal strength is high. So, we differentiate three kinds of vehicular node: 1) Vehicular GateWay (VGW) refers to the vehicle that has both IEEE 802.11p and EUTRAN LTE interfaces installed, enabling it to serve as an Internet mobile gateway. Its primary objective is to extend Internet connectivity to other vehicles by transmitting an advertisement message to the reversed direction of the route within the constricted proactive broadcast area; 2) Vehicular Gateway Candidate (VGC) is a vehicle with dual interfaces witch enters in 4G active region and its Receive Signal Strength (RSS) concerning the LTE base station (LTE-BST) LTE-RSS exceeds a certain threshold $RSS_{Th}$, this type of vehicle can be designated as VGW or relay in the network. 3) Ordinary vehicle (OV) is a vehicle with IEEE 802.11p capabilities that can play only the role of the relay to assist the vehicular source nodes to discover proactively the itinerary on the way to the VGWs. Vehicles with dual interfaces can be considered OVs when they are outside of the active region or their RSS drops below than certain threshold.

If one vehicle desires to connect to the Internet and did not hear the advertisement message or there is no entry corresponding to the one of VGW in its routing table, it broadcasts in the same manner a solicitation message to research reactively paths toward one of the available VGW in the respect of its reactive broadcast zone.

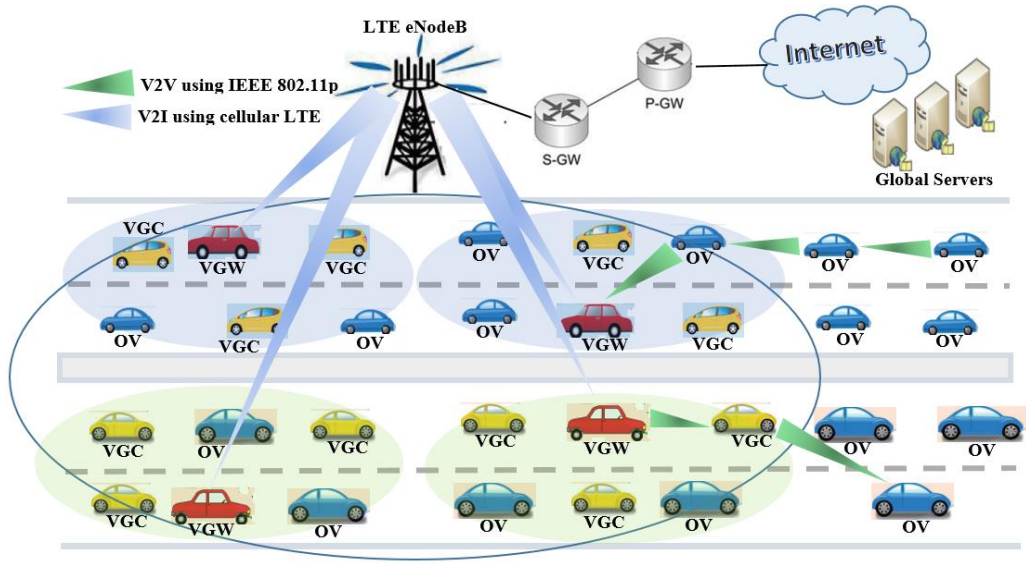

Figure 1. Designed VANET-LTE heterogeneous network architecture

## 3.2. Proactive Internet Mobile Gateway Discovery Approach

The Internet access for VANETs is generally provided via Internet mobile gateways that are positioned in the 4G active region of LTE BST. To start an Internet connection, in a first step these mobile gateways have to be discovered by the vehicles within the predefined zone. As known table-driven routing (proactive) approach minimizes discovery latency, whilst on-demand-based routing (reactive) approach reduces Control message overhead experienced by the network during route discovery [3, 10], the proposed protocol has envisaged a hybrid gateway discovery. In the proactive Internet mobile gateway discovery process, vehicles designated to act as mobile gateways are asked to broadcast periodically an Internet Mobile GateWay ADVertisement (IMGW_ADV) message within the restricted zone that can be determined using hop-count metric or indeed the time to live. The discovery of Internet mobile gateway procedure is carried out by spreading the IMGW_ADV messages hop by hop in a pre-determined area called the proactive broadcast zone. A message coming from the gateway VGW must not take place exterior of this area, and in that an efficient use of the network resources.

Regarding the broadcast of advertised messages, the proposed protocol used an efficient scheme based on contention-based forwarding approach [10],[13]. The scheme allows only some nodes in VANETs selected as potential relays to rebroadcast IMGW_ADV messages in predefined proactive broadcast zone. The objective is to reduce overhead, minimize delay and use efficiently the bandwidth, during the gateway discovery process. Each Internet mobile gateway advertises periodically the vehicles within predefined zone for its availability. The IMGW_ADV message comprises a set of information such as the IP address of VGW, IP address of next hop, sequence number, sender stability parameters (location, direction, and speed), number of hops, effective route stability and available buffer queue of route. VGW initially sets the relay IP address to their own IP address; the effective route stability to a wide value; available buffer queue of route to its current and the hop-count to 0. Table 1 gives the general format of IMGW_ADV message.

Table 1. IMGW_ADV message structure

| Field | Description |
|---|---|
| IMGW_IP | Address of source Internet mobile gateway |
| Relay_IP | Address of relay |
| Seq_Number | Sequence Number of message |
| Broadcaster_Location | Geographic position of the broadcaster (gateway or relay) |
| Broadcaster_Speed | Speed of the broadcaster |
| Broadcaster_Direction | Direction of the broadcaster |
| ERS | Effective route stability |
| ABQ_route | Available route buffer queue |
| Nbr_Hops | Number of hops |

The presented routing protocol consists of four principals axes: vehicular mobile gateway designation, vehicular relay selection, path establishment, and hand-off process.

## 3.3. Vehicular Internet Mobile Gateway Selection Scheme

As discussed previously, in our scenario, there are only some vehicles, are equipped with dual EUTRAN LTE and IEEE 802.11p interfaces. When it enters within the 4G activate region, each one of these mobile vehicles uses the LTE-RSS to determine whether it can act as Internet mobile gateway.

### 3.3.1. Gateway Selection Metrics

#### 3.3.1.1. LTE Link Stability

To elect more appropriate VGCs to act as VGWs in the network, the proposed IMGsdRP takes into consideration link stability with LTE base station as a primary gateway selection metric. To measure this metric, the link lifetime (LLT) is calculated using the following equation:

$$LLT_{VGC} = \frac{R_{4G} - D_{VGC}}{V_{VGC}} \quad (1)$$

Where, $R_{4G}$ is the coverage of VGCs in LTE network, $D_{VGC}$ is a distance between LTE-BST and vehicle VGC and $V_{VGC}$ is speed of vehicle VGC.

#### 3.3.1.2. Speed Variation Rate

In a vehicular environment where the mobility of nodes is fast and varied frequently, the selection of VGWs based only on the LLT metric seems insufficient. The persistent change in speed can cause link instability and channel fading. As consequence, the packet error can be increased and the probability that packets are received successfully can be reduced as well [14]. The purpose of incorporating this metric is to avoid designating vehicles as VGW that experience frequent changes in speed. In this protocol, a new factor that reflects the degree of speed change in the link between VGC and LTE BST through time is suggested. In this effect, for each VGC, speed samples $(V_0,V_1,…,V_{n-1})$ are recorded for each instant $t_0$, $t_1$, ..., $t_{n-1}$ in each time interval not exceeding $\tau$ seconds. Over time, we slide our sampling window to get the latest values on speed changes. The vehicle which has high chance to be designated as VGW is that has less variation in its speed. A relative standard deviation of VGC speed can be used to represent the degree of variation. If we compute the current mean of sampled data of speed V noted $\bar{V}$, we can easily calculate the relative standard deviation (RSD) to reflect the variation of speed in respect of average speed, we have :

$$RSD = \frac{\sigma}{\overline{V}} \tag{2}$$

Where, $0 < RSD \leq 1$ and $\sigma$ is a standard deviation calculated as:

$$\sigma = \sqrt{\frac{\sum_{k=0}^{n-1} \left( V_k - \overline{V} \right)^2)}{n}} \tag{3}$$

### 3.3.2. Vehicular Gateway Selection Algorithm

Using a Contention-Based-Broadcasting strategy, the proposed vehicular Internet mobile gateway selection mechanism may be able to specify some VGCs with high optimality to conduct as gateways to the Internet. After the periodic time interval expires, neighboring VGCs wait for a delay before broadcasting the message, if during this period a VGC has not received any message, it broadcasts and becomes a VGW. The others who hear this message cancel their timer without broadcasting and keep their status at VGC. The priority to broadcast advertisement message is defined as a weighted function denoted $W_{VGC}$, which incorporates metrics such as LLT and RSD. To be sure that a value of $W_{VGC}$ is between 0 and 1, the proposed metrics must be normalized. Let $\mu$ represent the weighted factor corresponding to LLT and inversely to RSD. $W_{VGC}$ can be computed as:

$$W_{VGC} = \mu . \frac{LLT_{VGC}}{LLT_{max}} + (1 - \mu) . (1 - RSD_{VGC}) \tag{4}$$

Where $\mu$ is in [0,1], $LLT_{max}$ is the maximum link life time. Let $t(.)$ be the waiting time of each VGC. Thus,

$$t(W_{VGC}) = T_{max} \times (1 - W_{VGC}) \tag{5}$$

Where, $T_{max}$ is a maximum waiting time for broadcasting.

Three parameters are taken into consideration when selecting the vehicular gateway namely, LTE-RSS, LLT and RSD. If the LTE-RSS is greater than a specific received signal strength threshold $RSS_{Th}$, the vehicle is declared as VGC. If LLT surpasses certain threshold $LLT_{Th}$ and the degree of fading stays under certain threshold $RSD_{Th}$, it will contend to be the Internet mobile gateway. The detail is given to Algorithm 1.

**Algorithm 1:** Distributed Vehicular Internet mobile gateway selection

```
begin
    if multiple interface vehicle enters the coverage area of LTE-BST then
        if LTE-RSS ≥ RSS_Th then
        |   status = VGC;
        end
    end
    foreach VGC at each end of periodic advertisement interval do
        Calculate RSD, LLT ;
        if (LLT ≥ LLT_TH && RSD ≤ RSD_TH) then
            Compute W_VGC ;
            Calculate a defer time t(.);
            Run advertisement timer;
            if Advertisement timer expires then
            |   status=VGW ;
            |   Broadcast IMGW_ADV message including current parameters;
            else
                if It receives advertised message then
                |   Cancel advertisement timer;
                |   No change in status;
                |   Execute algorithm 2 ;
                end
            end
        else
        |   No change in status;
        end
    end
end
```

## 3.4. Vehicular Relay Selection Scheme

### 3.4.1. Relay Selection Metrics

#### 3.4.1.1. Effective Link Stability

To evaluate the level of effective stability of a link, the vehicular relay selection scheme exploits a new metric called ELS (Effective Link Stability). This metric calculates the lifetime of the link to anticipate the period of its breakage. The metric is calculated based on speed, direction and position as well as channel fading-based transmission range.

We will start with estimating transmission range based on channel fading statistics on the link. Let P be the power received which is considered to be a random variable that behaves according to a Nakagami distribution under the fading channel model in vehicular environments [15]. As explained in [16], with power x, if we have received a signal for a given average power $\Omega$d at distance d, the probability density function (PDF) can be expressed as follows:

$$f_P(x) = \left(\frac{\Omega_d}{m}\right)^m \frac{x^{m-1}}{\Gamma(m)} \exp\left(-\frac{x^m}{\Omega_d}\right), x \geq 0. \tag{6}$$

$\Gamma(.)$ is the gamma function and the parameter $m$ is the fading intensity. Let $P_r$ be the probability where a carrier signal $x$ of the packet arrives with a received power exceeds a certain threshold $P_{th}$. The $P_r$ can be written as:

$$
\begin{aligned}
P_r\{P \geq P_{th}\} &= \int_{P_{th}}^{+\infty} f_P(x)dx \\
&= \int_{P_{th}}^{+\infty} \left(\frac{\Omega_d}{m}\right)^m \frac{x^{m-1}}{\Gamma(m)} \exp\left(-\frac{x^m}{\Omega_d}\right)dx
\end{aligned}
\tag{7}
$$

In the paper [16], if *m* is a natural number, the authors have demonstrated that the probability $P_r$, in discrete domain, can be reformulated as follow:

$$
P_r\{P \geq P_{th}\} = \sum_{k=0}^{m-1} \frac{1}{k!} \left(\frac{mP_{th}}{\Omega_d}\right)^k \exp\left(-\frac{mP_{th}}{\Omega_d}\right)
\tag{8}
$$

For reason of simplicity, the transmission power is assumed to be constant and identical for all vehicles. In the event of non-interference, the average received power $\Omega_d$ is supposed to follow the Friis model where path loss exponent equals 2:

$$
\Omega_d = KP_t d^{-2}
\tag{9}
$$

Thus, the signal received with power threshold $P_{th}$ should be, on average, detected at the maximum transmission range *R*:

$$
P_{th} = KP_t R^{-2}
\tag{10}
$$

where, $K=G_t G_r\lambda^2/(2\pi)^2 L$ is a constant value, $\lambda$ is the wavelength, $G_t$ and $G_r$ are respectively antenna gains of the transmitter and receiver, and *L* is the path loss factor, usually set to 1. We substitute each parameter in equation 8 and we finally get:

$$
P_r\{d \leqslant R\} = \exp\left(-m\left(\frac{d}{R}\right)^2\right) \times \sum_{k=0}^{m-1} \frac{1}{k!} \left(m\left(\frac{d}{R}\right)^2\right)^k
\tag{11}
$$

Note that $P_r$ can be calculated if the parameters m and *R* are available. The inter-vehicle distance *d* separated two vehicles is used to estimate the parameter *m* [17].

$$
m(d) = \begin{cases} 1 & d \geq 150m \\ 1.5 & 50m \leqslant d < 150m \\ 3 & d < 50m \end{cases}
\tag{12}
$$

The above equation 11 denotes that the probability where two nodes (vehicles) separated by distance *d* can stay connected with each other. For fading channel the average expected transmission range E[R] denoted $R_{exp}$ can be found from [18]:

$$
R_{exp} = E[R] = \int_0^{+\infty} P_r\{z \leqslant R\}dz
\tag{13}
$$

According to the *m* values, we have:

$$R_{exp} = \begin{cases} I = \int_0^{+\infty} \exp\left(-\left(\frac{z}{R}\right)^2\right) dz & m = 1 \\ \frac{1}{2}\left[\int_0^{+\infty} \exp\left(-2\left(\frac{z}{R}\right)^2\right)\left(1 + 2.\left(\frac{z}{R}\right)^2\right) dz + I\right] & m = 1.5 \\ \int_0^{+\infty} \exp\left(-3\left(\frac{z}{R}\right)^2\right)\left(1 + 3.\left(\frac{z}{R}\right)^2 + \frac{9}{2}.\left(\frac{z}{R}\right)^4\right) dz & m = 3 \end{cases}$$

(14)

Let *R* be the maximum wireless transmission range of the IEEE 802.11p interface of the two vehicles. By applying an integration by substitution method, the expression $R_{exp}$ can be simplified as follow:

$$R_{exp} = R.\left(1 - \varepsilon\right)$$

(15)

Where,

$$\varepsilon = 1 - \begin{cases} I = \int_0^{+\infty} \exp\left(-x\right)^2 dx & m = 1 \\ \frac{1}{2}\left[\frac{\sqrt{2}}{2}\int_0^{+\infty} \exp\left(-x^2\right)\left(1 + x^2\right) dx + I\right] & m = 1.5 \\ \frac{\sqrt{3}}{3}\int_0^{+\infty} \exp\left(-x^2\right)\left(1 + x^2 + \frac{1}{2}.x^4\right) dx & m = 3 \end{cases}$$

(16)

Knowing that:

$$I = \int_0^{+\infty} \exp\left(-x\right)^2 dx = \frac{\sqrt{\pi}}{2}$$

(17)

The final value of ε is:

$$\varepsilon = 1 - \begin{cases} \frac{\sqrt{\pi}}{2} & if \; m = 1 \\ \frac{\sqrt{\pi}}{16}.\left(3\sqrt{2} + 4\right) & if \; m = 1.5 \\ \frac{5\sqrt{3}\sqrt{\pi}}{16} & if \; m = 3 \end{cases}$$

(18)

The parameter ε represents the proportion of degradation of the maximum transmission range. It is used to know the fading conditions of the wireless channel in the current link.

The effective link stability (ELS) represents the link lifetime computed based on the mobility parameters, namely, position, speed, direction [10], as well as expected wireless communication transmission range of the vehicular node denoted $R_{exp}$. Let i and j be two vehicles which are positioned at $(x_i, y_i)$ and $(x_j, y_j)$ and moving with speed $v_i$ and $v_j$, in directions $\theta_i$, $\theta_j$ with respect to the x-axis, respectively. To estimate the value of $LET_{ij}$, we take advantage to the formula proposed in [19].

$$ELS_{ij} = \frac{\sqrt{(a^2 + c^2)R_{exp}^2 - (ad - bc)^2} - (ab + cd)}{a^2 + c^2}$$
$$a = v_i \cos\theta_i - v_j \cos\theta_j$$
$$b = x_i - x_j$$
$$c = v_i \sin\theta_i - v_j \sin\theta_j$$
$$d = y_i - y_j.$$

(19)

We define effective route stability *ERS*, the lifetime of a route, as the minimum *ELS* over all ℓ links in the path, such that:

$$ERS = \min_{j=0}^{\ell-1} ELS_j$$

(20)

### 3.4.1.2. Effective Distance Rate

By using ERS as a routing metric, we will be sure that the selected route is the most stable and the least faded. This will definitely enhance network performance, especially in terms of data throughput. However, a most stable path may have a higher hop count. When packet is relayed through more intermediate nodes, it will increase medium access contention, interference, congestion, and packet collisions. Therefore, hop-count should also be taken into account when selecting a suitable path. For this reason, one more parameter named EDR (Effective Distance Rate) will be included. EDR characterizes the progress rate of the announcement message towards the last vehicle of the zone of interest in the opposite movement direction. Therefore, the farthest vehicle, which received the message successfully, will be prioritized to rebroadcast message. As result, incorporating this metric will allow vehicles to choice the paths with lower hop-count. The EDR is defined as:

$$EDR_{ij} = \frac{\min(d_{ij}, R_{exp})}{R_{exp}}$$

(21)

$d_{ij}$ is the distance between the previous transmitting vehicle i and the current receiving vehicle j. $R_{exp}$ is the effective communication range of the vehicular nodes.

### 3.4.1.3. Available Buffer Queue

In VANET, especially in high vehicle density scenarios, a single vehicle can serve as the next hop for multiples vehicles. When the number of demands is over its capacity of service, the vehicle may be encumbered and possibly lose packets, and an increase in delay and MAC re-transmissions will be observed. The aim is to avoid vehicles acting as relays do not become overloading in the network. For this purpose, another routing metric has been considered in the relay selection. This metric named Available Buffer Queue (ABQ) proposed in [20]. Its purpose is to determine the amount of data traffic that is directed towards vehicle. Thus, taking this metric into account will be useful to prevent bottlenecks in the network, choose a less congested route, and reduce efficiently data packet loss. At one vehicle v in the network with a maximum queue length $L_{max}$, its ABQ can be computed using the following formula :

$$ABQ_v = 1 - \frac{L}{L_{max}}$$

(22)

Where L is the current number of data packets queued in vehicle buffer.
The available buffer queue of a route is the sum of $ABQ_j$ at any vehicle j, including relays and the mobile gateway, in that route. The overall ABQ of route ($ABQ_{route}$) is computed as

$$ABQ_{route} = \sum ABQ_j$$

(23)

### 3.4.1.4. Relay Selection and Rebroadcasting

Once one vehicle is designated to behave as VGW, it diffuses an IMGW_ADV message in the opposite direction of movement inside the proactive broadcast zone. In our protocol IMGsdRP, only some potential vehicles that are selected as relays for retransmitting such messages to the neighboring vehicles in this zone. As previously mentioned, the selection of the relay is realized using contention-based forwarding (CBF) mechanism [13], following the same process in [10]. Here, the best relay should be the furthest vehicle that has the most stable and highest available buffer queue with the sender. Thus, the contention is based on the three parameters cited early, such as effective link stability (ELS), effective distance rate (EDR) and queue available buffer (ABQ). Thus, we have modified the CBF parameter with a new function that verifies our preconditions. To form this multi-metric function which combines the impact of the three metrics in varying proportions, the proposed protocol takes advantage to the weighted mean technique. Since the values used in the proposed relay selection mechanism must be between 0 and 1, so before weight calculation, we firstly have to scale the ELS metric to bring its value into normalized and non-dimensional value. However, the normalization of parameter EDR and ABQ is not required.

Concerning ELS metric, we propose an effective stability function denoted $S_e$, which depends on the effective stability of the link. We consider an exponential function cited in [10], which meets the accorded criteria. Note the effective stability function:

$$S_e = 1 - exp\left(-\frac{ELS}{A}\right)$$

(24)

where A is a constant indicating the growth rate of the function and we have :

$$\lim_{ELS \to +\infty} S_e = 1 \ \ and \ \ \lim_{ELS \to 0} S_e = 0$$

(25)

i.e, the greater value of ELS is, the nearer the result of this function $S_e$ is to 1, against, the minor the ELS is, the function $S_e$ gets closer to 0. The weight of each vehicle is computed basing on the following function noted W.

$$W = \alpha \times S_e + \beta \times EDR + \gamma \times ABQ$$

(26)

$\alpha$, $\beta$ and $\gamma$ are the are the weighting factors, where $0 \leq \alpha$, $\beta$, $\gamma \leq 1$, used in order to give more weight to one metric than others, with $\alpha + \beta + \gamma = 1$.

As detailed in Algorithm 2, upon reception of IGMW_ADV message, in the opposite direction inside proactive zone, the vehicle prepares for a while of time to spread the message. This waiting time is computed employing the weighted function (W). For the contention, the timer is set as follows:

$$t(W) = T \times (1 - W)$$

(27)

Where $0 \leq W \leq 1$, $0 \leq t(W) \leq T$, and T is the maximum forwarding delay. During this time, if the vehicle receives a message similar to the previous one (meaning the message has been broadcast yet), it will deactivate the timer and reject the two advertisement packets. Else, the vehicle rebroadcasts the packet with new information after its timer is over, which will stop other vehicles from broadcasting.

**Algorithm 2:** Algorithm of Relay Selection

**begin**
   **Upon reception of IMGW_ADV packet;**
   **if** *vehicle is VGW* **then**
     |   Ignore received packet and return;
   **end**
   **if** *The IMGW_ADV packet is received for first time && the vehicle (OV or VGC) moves in the*
   *same direction && the veichle is behind* **then**
     |   Calculate $S_e$, $ERD$ and $ABQ$ ;
     |   Compute $W$ ;
     |   Set a defer time according to the fomula 27;
     |   Launch timer;
   **else**
     |   Call off timer;
     |   Ignore the two IMGW_ADV packets;
   **end**
   **Upon expiration of the waiting time ;**
   Modify the previous mobility and routing information with the current in the IMGW_ADV packet;
   Rebroadcast IMGW_ADV message;
**end**

## 3.5. Route to the Vehicular Mobile Gateway Discovery

### 3.5.1. Proactive Approach

Upon reception of IMGW_ADV message, vehicles will have one or more itineraries to the Internet gateways in the routing table. Each IEEE 802.11p interface of vehicular node conserves a routing table to store, update and delete routes to the mobile gateways. As shown in Table 2, the routing table has the following fields: VGW address, next hop address, sequence number, effective route stability, and number of hops. Vehicles calculate the new ERS, new ABQ$_{route}$ and the other information are extracted from the received message.

Table 2. Routing table structure

| IMGW_IP | NextHop_IP | SeqNo | ERS(s) | Nbr_Hops | ABQ$_{route}$ |
|---------|------------|-------|--------|----------|---------------|
| g | n | s | e | h | q |

On receiving the IMGW_ADV packet in the proactive area, if the vehicle does not have any entry corresponding to the gateway which is the originator of the message, it updates the routing table by adding a new entry. Otherwise, if the routing table comprises an entry matching the IP address of the mobile gateway, and the message has arrived with a higher sequence number, then the entry is updated with the current values involved in the message. Nevertheless, if the sequence number of the received message is equal to that of the existing routing entry in the cache, if the expired time of entry is less than ERS included in the message, the vehicle updates the corresponding entry in its routing table from the information of the IMGW_ADV packet. In the case of equality, the update of the entry will occur if the message arrives with a lower hop-count value. If the number of hops is the same, the optimal route is the one with the maximum ABQ. After its lifetime expires, the route is deleted from the routing table.

**3.5.2. Reactive Approach**

When one source vehicle needs to connect to the Internet, and it does not hear any announcement within the interest zone, a reactive discovery mechanism is triggered. In this context, an Internet Mobile Gateway solicitation message (IMGW_SOL) is broadcasted by exactly the same mechanism as IMGW_ADV until it receives by a VGW or any vehicular node intermediate that its routing table comprises an itinerary towards one VGW, we called this vehicle awareness vehicle (AV). Once a vehicle AV or VGW receives IMGW_SOL message in the reactive broadcast zone of the source (Nbr_Hops<k) [21], after being victorious in contention, it sends a unicast (IMGW_ADV) message to the IMGW\_SOL message originated source vehicle. Upon reception of unicast IMGW_ADV, each intermediate vehicle computes ELS and ABQ parameters and compares them with ERS and ABQ$_{route}$ integrated into the message respectively. The IMGW_ADV message, as well as the routing table, will be updated with the minimum values and the number of hops is incremented by 1.  Upon receipt of the unicast IMGW_ADV packet as a reply, the vehicular source upgrades its routing table. Route stability, hop count and available buffer queue are all metrics used for route selection. If multiple responses are received from one vehicular source, route selection is applied in that order as well as in a proactive approach.

## 3.6. Handover Process in VANETs

For guaranteeing seamless transfer connectivity of one vehicle in the network from one VGW to another, an efficient handover process is required.  Handover connections to the next VGW occur seamlessly and before the current connection ends. To set up the connection, the vehicles record constantly the optimal routes in their routing table. The process of handover can be triggered explicitly when the current serving VGW loses its optimality, it sends an unicast notification message toward each source using it as gateway. Implicitly, to maintain the connection, at the end of a certain time called critical time, the vehicles cheek their routing tables in order to discover the most optimum route to the new VGW, and start to hand connection over. The critical time ($T_c$), is the moment when the vehicles take decision that the actual route is expiring and the process of handover should begin [10]. Thus, the $T_c$ is defined as follows:

$$T_c = ERS - \delta$$

(28)

Where $\delta$ is a back-off value that allows vehicles to take correctly action in a timely manner. In the present protocol, we have based on a route discovery latency to put value of $\delta$. The more details for hand-off mechanism are given to Algorithm 3.

---

**Algorithm 3:** Vehicular Gateway Handover Mechanism

**begin**
  **foreach** *VGW vehicle in VANET* **do**
    **if** *LTE-RSS $< RSS_{Th}$* **then**
      | Send an unicast Notify message to each source using it as Internet gateway;
    **else**
      **if** $(LLT_{VGW} < LLT_{Th} \,||\, RSD > RSD_{TH})$ **then**
        | Send an unicast Notify message to each source using it as Internet gateway;
      **end**
    **end**
  **end**
  **foreach** *Active vehicle source in VANET* **do**
    **On Adding or updating route entry;**
    set handover timer to critical time $T_c$;
    run timer ;
    **On receiving Notify message from VGW;**
    **if** *Handover timer status is PENDING* **then**
      Cancel Handover timer ;
      **if** *An optimal route exists to an other VGW in routing table* **then**
        handoff to the next VGW;
        Send Thanks message to the old VGW;
      **else**
        | Broadcast IMGW_SOL message to find reactively new potential VGW;
      **end**
    **end**
    **On the critical time $T_c$ expire;**
    **if** *An optimal route exists to an other VGW in routing table* **then**
      handoff to the next VGW;
      Send Thanks message to the old VGW;
    **else**
      | Broadcast IMGW_SOL message to find reactivelly new potential VGW;
    **end**
    **On the receiving replay of IMGW_SOL message;**
    handoff to the new elected VGW;
    Send Thanks message to the old VGW;
  **end**
**end**

---

# 4. RESULTS AND DISCUSSION

The simulation environment that we consider to evaluate the performance and efficiency of the proposed protocol is composed of two simulators. We chose NS2 (Network simulator 2) [22] as a network simulator because it has a set of diversified network components that allow us to conduct different conclusive studies depending on the protocols to be analyzed. The use of NS2 is popular in the field of studies related to wireless network technologies as well as the development of ad hoc routing protocols. NS2 supports IEE 802.11p in the form of two modules WirelessPhyExt and 802_11Ext which allow us to carry out realistic studies on VANETs networks, while LTE-patch is used to simulate LTE network. In order to test our protocol in a realistic environment, we use SUMO (Simulation of Urban MObility) [23] as a traffic simulator to generate the mobility model. It is frequently used for studies related to traffic in a road network and is highly portable. Parameters of simulation environment are all given to Table 3 and Table 4.

Table 3. Mobility features

| Parameter | Value |
|-----------|-------|
| Area | 8000x100 m² |
| Mobility model | Manhattan mobility model |
| Number of lanes | 2 for each direction |
| Maximum speed | 30 m/s |
| Number of vehicles | 50 |
| Number of dual interface vehicles | 5,10, 20, 30, 40 |
| Simulation time | 300 s |

Table 4. Integrated network parameters

| Parameter | Value |
|-----------|-------|
| Channel | Channel/WirelessChannel |
| Network Interface | Phy/WirelessPhyExt |
| MAC | Mac/802\_11Ext |
| Antenna Type | Antenna/OmniAntenna |
| Interface queue | Queue/DropTail/PriQueue |
| Data rate& | 6 Mbps |
| Interface queue size | 20 |
| IEEE 802.11p transmission range R | 250m, 300m, 350m, 400m, 450m |
| Routing protocols | IMGsdRP, RTMGwS |
| Traffic type | CBR |
| Packet sending interval | 0.1 s |
| Packet size | 1000 bytes |
| LTE | ns-2.34-LTE & multi-interface patches |
| Transmission range of eNodeB | 7 km |

The recent RTMGwS routing protocol proposed in [9] is used to compare its performance with our protocol IMGsdRP in two cases: IMGsdRP-ETR and IMGsdRP-MTR which are based on expected transmission range $R_{exp}$ and maximum transmission range R to calculate ELS and EDR metrics respectively.

We have selected the metrics that we consider the most significant to evaluate the performance of the proposed routing protocol, namely:

- The Packets Delivery Ratio is the ratio of the number of packets routed successfully to the total number of packets transmitted by the source. It characterizes transmission reliability.
- The End to End packet Delay is the time between the moment a packet is sent by the source transport layer and the moment the packet is received by the destination transport layer. This metric represents the efficiency of the protocol in terms of response time and in terms of the discovery optimal paths.
- The Normalized Routing Overhead represents the number of control packets sent for each data packet received. This metric allows us to evaluate for each protocol, the overhead caused for sending control packets.

## 4.1. Effect of Number of Multiple Interfaces Vehicles in the Network

In this simulation, the number of VGCs was varied in the network from 5 to 40, to see the impact on the performance of protocols. The number of sources is taken to 5 vehicles which are randomly designed to send CBR data at a rate of 10 packets/s to the server node. The effectiveness of the three protocols in terms of packet delivery ratio, average end-to-end delay, and normalized routing overhead, varying the number of VGCs is illustrated in Figures 2, 3 and 4 respectively.

Figure 2 depicts the result of the average packet delivery ratio for the three protocols. From the curves in the figure, it can be seen that IMGsdRP in its two cases outperforms the RTMGwS routing protocol for different numbers of VGCs. For both protocols, IMGsdRP-MTR and RTMGwS, the result is considered good and close with a slight superiority of the former. This is due to the fact that in both protocols, the vehicles constantly keep in their routing tables the optimal and recent paths that can be adopted as backup paths. Moreover, the efficient distributed manner used to broadcast advertisement messages in the network consumes a little bandwidth and large remains for transferring data. Slight-out performance is given by used metrics and the improvement of both mechanisms gateway selection and handover. As shown in the figure, the proposed routing protocol IMGsdRP-ETR yields 3.28\% improvement in increasing packet delivery ratio compared with IMGsdRP-MTR, and 12.83\% improvement compared to RTMGwS. the main cause behind the protocol IMGsdRP-ETR gains better than the others is that the chosen routes are those having the longer lifetime which is measured effectively taking into consideration the wireless channel fading, therefore, the route with the high probability of packet reception is selected. In addition, IMGsdRP selects paths having a fewer number of hops based on EDR. This reduces congestion, contention, and interference, and as consequence packets received successfully are increased.

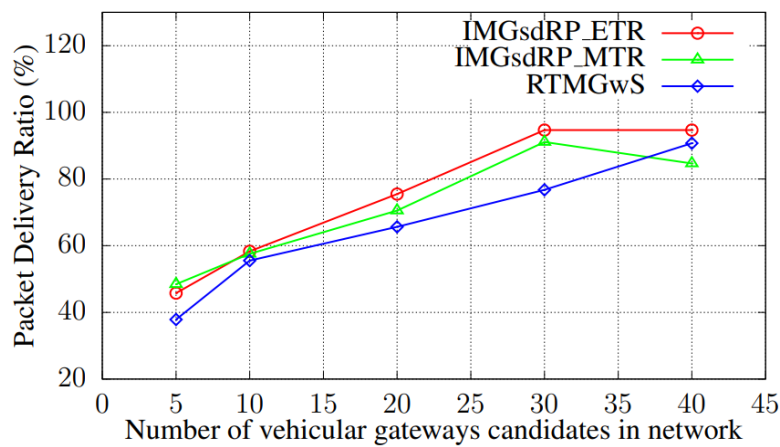

Figure 2. Packet delivery ratio comparison under different number of vehicular gateway candidates in the network

Figure 3 depicts the average end-to-end delay experienced by data packets for varying gateway candidate vehicle density in the network. As predicted, for all of the routing protocols, packet delay slightly decreases. The cause is whenever the number of VGCs increases, the time of route discovery is reduced. On the whole, the protocol IMGsdRP-ETR presents end-to-end delay slightly higher than others. The reason behind this slight improvement equals to 6.02\% over IMGsdRP-MTR and 5.95\% over RTMGwS is that in IMGsdRP-ETR, the vehicles have always in their routing table, the updated optimal routes selected based on the effective route stability metric, and no spent time for researching another route in case of fainting of the current route. Moreover, the shorter route selected helps to minimize the delay experienced by the packets from sources to the VGWs.

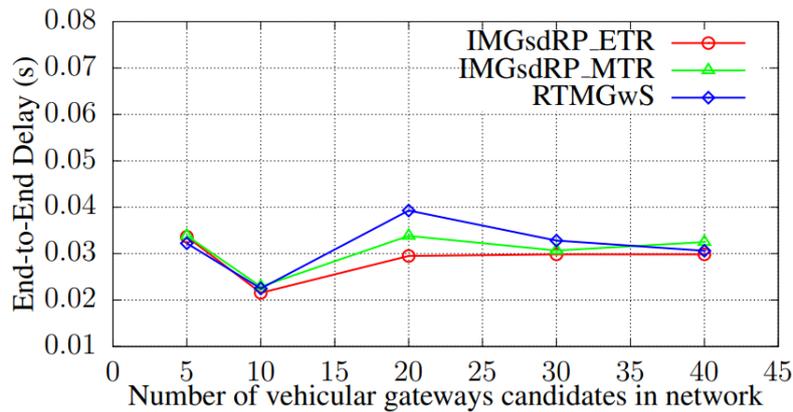

Figure 3. End-to-End delay comparison under different number of vehicular gateway candidates in the network

The result of the normalized routing protocol comparison is illustrated in Figure.4. It depicts that by varying the number of VGCs, the overhead rises appropriately, for the three protocols. In both protocols, IMGsdRP and RTMGwS, more VGCs in the network mean the number of vehicles that may become an Internet mobile gateway increases, and the announcement packet has to be diffused to more vehicles than the previous, as consequence, more overhead will be generated in the network. As shown in the figure IMGsdRP-ETR presents on average 5.6\% and 44.3\% improvement in reducing routing overhead compared to IMGsdRP-MTR and RTMGwS respectively. The first result is due to the integration of metrics that are based on the expected transmission range in the route selection, while the second, is thanks to the vehicular gateway selection scheme that aims to minimize the number of VGWs in the network.

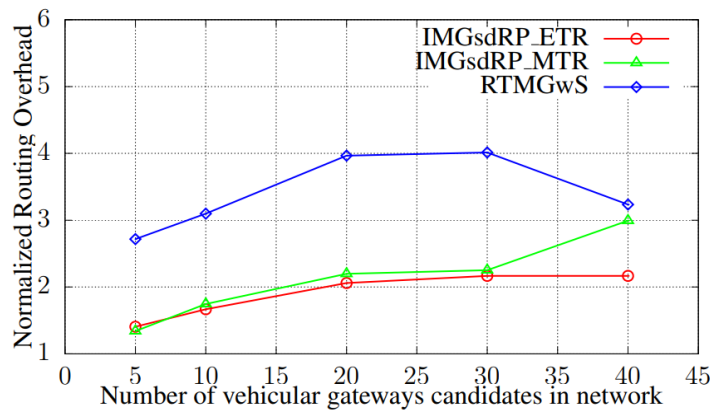

Figure 4. Routing overhead comparison under different number of vehicular gateway candidates in the network

## 4.2. Effect of Maximum Transmission Range

To analysis the effect of IEEE 802.11p maximum wireless transmission range on the performance of the routing protocols, the number of VGCs is taken as 20 and the number of vehicular sources as 5 selected randomly, the CBR sending rate is 10 packets/s and maximum vehicle speed 30 m/s. As illustrated in Figures 5, 6 and 7, various IEEE 802.11p wireless transmission ranges are explored to analyze the impact on the effectiveness of the three simulated protocols. As shown in the figure, all routing protocols show that there is not a significant variation in the network performance with the increase of the maximum transmission range.

The packet delivery ratio by the three protocols for different values of IEEE 802.11p maximum transmission range is plotted in Figure.5. As demonstrated in the figure, the routing protocol IMGsdRP-MTR outperforms the protocol RTMGwS. This is because, in IMGsdRP-MTR, the more potential VGCs in terms of stability and channel quality with LTE-BST are those elected to be Internet gateways. This leads to reduce vehicular gateway handoffs and consequently packet loss decreases. However, the proposed protocol IMGsdRP-ETR gives better results than the two other protocols IMGsdRP-MTR and RTMGwS. The leading reason beyond is that the routing metrics used in IMGsdRP-ETR are estimated not only based on mobility information but also integrates the impact of wireless channel quality in term of the degree of fading. Thus, the frequent disconnections and route failures decrease in the network. IMGsdRP-ETR shows a 3.38\% improvement in terms of packet delivery ratio over IMGsdRP-MTR and a 13.6\% improvement over RTMGwS.

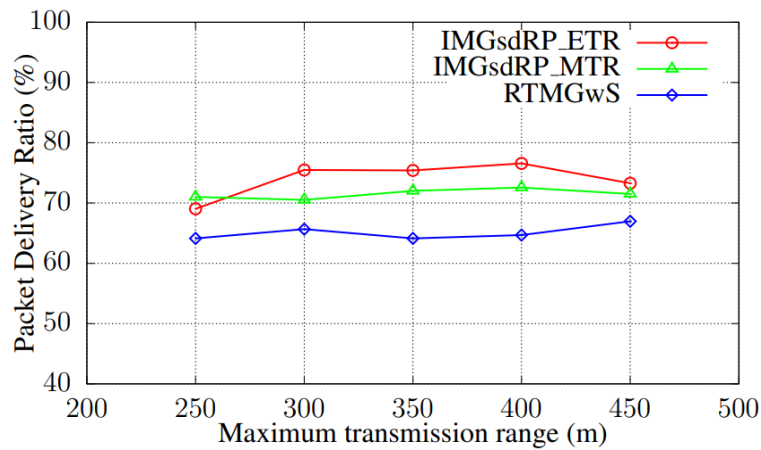

Figure 5. Packet delivery ratio comparison under different comparison under different maximum IEEE 802.11p wireless transmission range

The pace of variation in the average end-to-end delay according to IEEE 802.11p maximum transmission range of IMGsdRP-ETR, IMGsdRP-MTR and RTMGwS are shown in Figure 6. As shown in the figure, the IMGsdRP-ETR always outperforms the other two routing protocols, whereas, it exhibits approximately 7.56\% and 15.62\% decrease in average packet delay compared to IMGsdRP-MTR and RTMGwS respectively. RTMGwS introduces weak results, because the protocol is based on the hop count counter to select the routes, then sometimes routes can contain a large number of hops, which can may experience a significant contention delay at the MAC layer. Under high coverage, route with higher hop count is fragile because the fast mobility and wireless channel conditions might cause the route to failure.

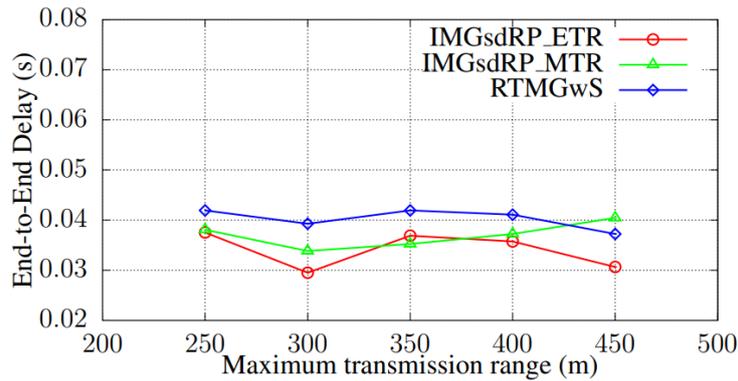

Figure 6. End-to-End delay comparison under different comparison under different maximum IEEE 802.11p wireless transmission range

Figure.7 shows the routing overhead generated by IMGsdRP-ETR, IMGsdRP-MTR and RTMGwS when varying the IEEE 802.11p maximum wireless transmission range in the network. As shown in the figure, the proposed protocol IMGsdRP outperforms significantly RTMGwS. This scenario is due to that when the maximum transmission range increase, the number of neighboring VGCs increases, and so, our protocol used an efficient way based on contention between several neighboring VGCs to select one of them for broadcasting advertisement messages. Thus our protocol minimizes the overhead required to discover the routes and selects more reliable routes from sources to the mobile gateway. IMGsdRP-ETR presents 3.28\% and 47.44\% improvement in reducing generated overhead over both protocols IMGsdRP-MTR and RTMGwS respectively.

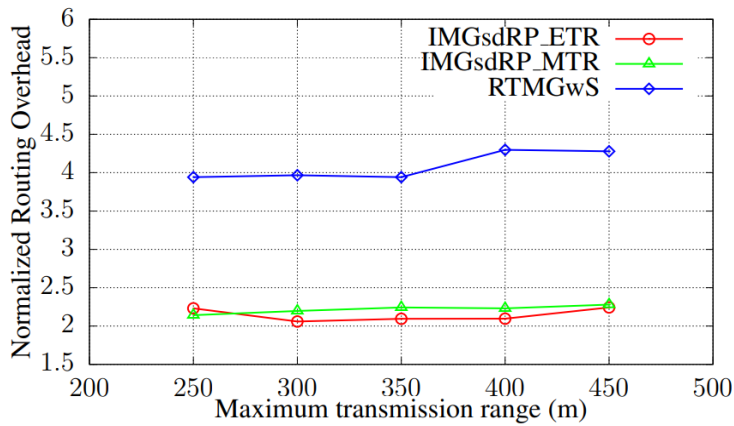

Figure 7. Routing overhead comparison under different comparison under different maximum IEEE 802.11p wireless transmission range

## 5. CONCLUSIONS

In this paper, a new routing protocol is introduced to connect moving vehicles to the Internet through some specific vehicles equipped with dual interfaces (i.e public transport vehicles) which are selected to serve as mobile gateways in an integrated VANET-LTE heterogeneous network. The proposed protocol integrates an efficient mechanism for rebroadcasting advertisement and solicitation messages to discover the gateways and establish most optimum routes from/to them. Received signal strength and degree of speed variation are the metrics used for designating optimal gateways in the network. However, mobility parameters and fading statistics based stability, inter-vehicular distance and buffer queue availability are the main metrics taken into consideration for discovering and selecting route to the vehicle gateways. Simulation results show that our proposed protocol improves network performance in terms of packet delivery ratio, packet delay, and overhead compared to existing protocols.

## AUTHORS

**Driss ABADA** is an associate professor at faculty of science, Chouaib Doukkali University, Morocco, since me 2019. He received his PhD degree in Computer Science with specialization Wireless Networks. His research interests are in wireless ad-hoc, Internet of things and 5G. Particularly, he works on wireless routing, routing protocols in Internet of things, inter-vehicular communications and security, Quality of service, and MAC layer performance evaluation. He can be contacted at email: abada.d@ucd.ac.ma.

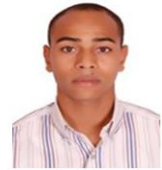

**Rachid ADRDOR** is currently a member of the Laboratory of Computer Information Systems and Vision of the Faculty of Science, Ibn Zohr University, Agadir, Morocco. He received the Ph.D. degree in Computer Science at the Faculty of Science of the same university. His research interests include Applied Artificial Intelligence, in particular, Distributed Constraint Optimization Problems (DCOPs) and Soft Arc Consistency.

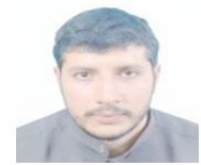

**Omar BOUTKHOUM** is an associate Professor at computer Science department in the Faculty of Sciences of Chouaib Doukkali University, EL Jadida, Morocco. He received his PhD degree in Computer Science from the Faculty of Sciences and Techniques of Caddi Ayyad Univer- sity, Marrakesh in 2017. His research interests are in the application of decision support systems and Blockchain technology to sustainable supply chain management. He can be contacted at email: boutkhoum.o@ucd.ac.ma.

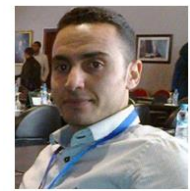